\documentclass[aps,pre,twocolumn]{revtex4}

\usepackage{amsmath,bm,epsfig}

  \usepackage[dvips]{color} 
  \usepackage{ulem}         
\newcommand{\tcblk}[1]{\textcolor{black}{#1}}

  \newcommand{\be}{\begin{equation}}
  \newcommand{\ee}{\end{equation}}
  \newcommand{\ba}{\begin{array}}
  \newcommand{\ea}{\end{array}}

\def \ed {\end{document}}
\def\Fbox#1{\vskip1ex\hbox to 8.5cm{\hfil\fboxsep0.3cm\fbox{%
  \parbox{8.0cm}{#1}}\hfil}\vskip1ex\noindent}  
\let \nn  \nonumber
\newcommand{\br}{\\ \nn}

\let\*\cdot
\def\<{\left\langle} \def\>{\right\rangle} \def\({\left(} \def\){\right)}
\let\p\partial \let\~\widetilde \let\^\widehat \def\ort#1{\^{\bf{#1}}}
 \def\x{\ort x} 
  \def\1{\bm1} 
 
\newcommand{\B}[1]{{\bm{#1}}}
\newcommand{\C}[1]{{\mathcal{#1}}}    
\def\BE{\begin{equation}}\def\EE{\end{equation}}
\def\BEA{\begin{eqnarray}}\def\EEA{\end{eqnarray}}
\def\BSE{\begin{subequations}}\def\ESE{\end{subequations}}

\renewcommand{\sb}[1]{_{\text {#1}}}  

\newcommand{\eq}[1]{(\ref{#1})}
\newcommand{\Eq}[1]{Eq.~(\ref{#1})}
\newcommand{\Eqs}[1]{Eqs.~(\ref{#1})}
\newcommand{\Fig}[1]{Fig.~\ref{#1}}

\renewcommand{\a}{\alpha}\renewcommand{\b}{\beta}

\def\k{\kappa}

\let\p\partial 
\def\x{\ort x}

 \def \vK {von-K\'arm\'an~}

\let \= \equiv

 \let\*\cdot
 \def\sb#1{_{\rm{#1}}}

 \def\({\left(} \def\){\right)}
 \def \[ {\left [} \def \] {\right ]}
   
    \let\^\widehat
  \let\-\overline

  \def\Eq#1{Eq.~\eq{#1}}
     \def\<{\left\langle}
   \def\>{\right\rangle}
   \def\ort#1{\^{\bf{#1}}}

   \let\~\widetilde

 \def \vK {von-K\'arm\'an~}

\def \Rt {\mbox{Re}_\tau}

\begin{document}

\title{ Analytical Model of the Time Developing
Turbulent Boundary Layer}
\author{Victor S. L'vov\footnote{Correspondence author, email:
 Victor.Lvov@Weizmann.ac.il} and  Anna Pomyalov}
\affiliation{Department of Chemical Physics, The Weizmann Institute
of Science, Rehovot 76100, Israel}
\author{Antonino Ferrante\footnote{Currently at the Graduate
Aeronautical Laboratories, California Institute of Technology,
Pasadena, CA 91125} and Said Elghobashi} \affiliation{Dept of
Mechanical and Aerospace Engineering, University of California,
Irvine, CA 92697, USA}
\begin{abstract}
We present an analytical model for the time-developing turbulent
boundary layer (TD-TBL) over a flat plate. The model provides
explicit formulae for the temporal behavior of the wall-shear
stress and both the temporal and spatial distributions of the mean
streamwise velocity, the turbulence   kinetic  energy and Reynolds
shear stress. The resulting profiles are in good agreement with
the DNS results of spatially-developing turbulent boundary layers
at momentum  thickness Reynolds number equal to 1430 and
2900~\cite{S1,S2,S3}. Our analytical model is, to the best of our
knowledge, the {\sl first} of its kind for TD-TBL.
\end{abstract} \maketitle

 \noindent {\bf Formulation of the problem.}
In this paper we derive an analytical model for the
time-developing turbulent boundary layer (TD-TBL) over a flat
plate. We consider a flat plate ($z=0$) submerged in an
incompressible viscous fluid at rest for time $t<0$. At time $t=0$
the fluid moves as a whole in the $x$-direction with velocity
$V_\infty$. This motion creates a boundary layer near the plate.
We assume that this boundary layer is turbulent. Note that in the
case of  TD-TBL all statistical characteristics  of the flow
depend only on the time $t$ and distance $z$ from the wall [e.g.
the mean streamwise velocity is $ V(z,t)$. \tcblk{ In contrast, in
a spatially-developing TBL (SD-TBL) the statistical
characteristics depend on two spacial directions, $z$ and $x$, and
the mean velocity has also $V_z$ component normal to the wall.
Nevertheless in the limit of large Reynolds number both the TD-TBL
and SD-TBL
 become asymptotically
equivalent (with the replacement $x\leftrightarrow V_\infty
t$)~\cite{07LPR}. Therefore it is reasonable to first consider the
simpler
  simple case of TD-TBL.}\\

\noindent \textbf{Definitions and Model Equations}. We start from
the Navier-Stokes equations for an incompressible fluid. The
velocity $\bm U(\bm r,t)$ is decomposed into the sum of its mean
value $\B V(z,t) \equiv \langle \bm U(\bm r,t) \rangle$ and a
fluctuating part $\B u(\B r,t)$, 
 $\bm U(\bm r,t) = \x {V}(z,t)  + \bm u(\bm r,t)$
Here $\x$ is the unit vector in $x$-direction,  $\B r=\{x\,, y\,,
z\}$ is three dimensional coordinate, and $\<\dots \>$ denotes
averaging in time and in the span-wise direction $y$.

The three main quantities in the model are the mean shear $S(z,t)$,
the tangential Reynolds stress $\tau(z,t) $ and the turbulence
kinetic energy per unit mass $K(z,t)$, defined as: \BE\label{defs}
 S(z,t) \equiv  \frac{\p V }{\p z }\,,\
 \tau(z,t)  \equiv   -  \langle u_xu_z\rangle \,,\
 \ K(z,t)  =  \frac12 \,  \langle |\bm u|^2\rangle\  .
\EE The mean momentum equation ( e.g.  Eq. (4.12) in \cite{Pope})
after integration over $z$ has the form:
\BE \label{ME1} 
 \nu S(z,t)  + \tau(z,t)=  \tau_*(t)+ \frac{\partial }{\partial  t} \int _0^z
 V (z',t)\, d z'\ .
 \EE Here $\nu$ is the kinematic viscosity and the  right hand
side (RHS)  is momentum flux toward the wall. The integration
constant $\tau_*(t)=\nu S(0,t)$ is the wall shear stress.

The turbulence kinetic energy conservation equation for a TD-TBL
(see e.g. Eq.
  (5.132) in \cite{Pope}) can be written as:
 \begin{equation}  \label{ME2} {\p K(z,t) }/{\p t}
+ {\C E(z,t)}+ \B \nabla\cdot \B T(z,t) = \tau(z,t) \,S(z,t)\ .
\end{equation}
The RHS of this equation represents the production of turbulence
kinetic energy by the mean shear. The two terms in the  left hand
side (LHS) : the rate of energy dissipation, $\C E $, and spatial
energy flux, $\B T $, require modeling via $S$, $\tau$ and $K$.

Equations~\eq{ME1} and \eq{ME2} for $S$ and $K$ are exact. In order
to solve them, we need to add a third equation for $\tau(z,t )$
  and model  both $\C E ( z ,t )$  and $\B T(z,t )$. It is reasonable
to assume that in the log-layer (region where the \vK log-law holds)
the shear and normal Reynolds stresses have the same $z$ and $t$
dependence, thus their ratio is constant:  $ \tau (z,t ) \big /K
(z,t )  = c^2$.

As suggested by~\cite{Pope}   we write the rate of turbulence
kinetic energy dissipation as
 $
\C E(z,t) =  b\,  [K(z,t)]^{3/2} \big / z$,  where  $b$ is a
positive constant and the spatial energy flux in the $z$ direction
as $
 T (z,t)= -   D(z,t)\,
 {\partial K(z,t)}/{\partial z}$, where $D(z,t)= d\,  z \, \sqrt{K(z,t)} $   is
the turbulence diffusivity and   $d$ is a positive coefficient.

Finally, we summarize the equations of the present  model
[Minimalist Model {(Min-Model)}]  for TD-TBL as
\BSE\label{MM}  \BEA   \label{MM1b}
\nu S(z,t)+  \tau(z,t) &=&  \tau_*(t)+  \int  _0^z d
z'\frac{\partial }{\partial  t}  V (z',t)\,,\br
  \Big[ \frac{\p}{\p t}+\frac{b
  \sqrt{ K (z,t) }}{z} &\!\!\!\! -\!\! \!\!&  d\, \frac{\p}{\p
z}z \sqrt {K (z,t) } \frac{\p}{\p z} \Big]  K (z,t) \\ &  =&
\tau(z,t) S(z,t)\,,\label{MM1c}\\
\label{MM1a} 
   \tau(z,t) &=&    c^2 K(z,t)\ .  \EEA 
\ESE 
The boundary conditions for \Eqs{MM} at the wall ($z=0$) and at
the edge of the TBL ($z=\C Z(t)$) are:
 \BSE\label{BK}\BEA\label{BKa} 
&& \hskip -1cm \mbox{at the wall}\ z=0: \   V(0,t)=0\,, \quad K(0,t)=0\,, \\
&& \hskip -1cm \mbox{at }\  z={\cal Z}(t):    \   V(\C Z,t) =
V_\infty\,,\  K({\cal Z}(t),t)=0\ .\label{BKb} \EEA\ESE  
The numerical values of the model coefficients  $b$ and $c$ are
prescribed to be $ 0.34$ and $ 0.53$ respectively to obtain
results in agreement with experiments ( e.g.~\cite{Pope}), and
with DNS data (see Fig. 3 in \cite{05LPT}). The third coefficient
$d$  is fixed equal to $0.07$ in order to match the role of the
turbulent diffusivity in fully-developed turbulent channel
flows~\cite{channel}.

We normalize the variables of the TD-TBL in "wall units" using the
friction velocity $u_*(t)$,   viscous length-scale $\ell_*(t)$,
and viscous time-scale  $t_*(t)$ defined as
\BE  \label{wallA} 
u_*\=\sqrt{  \tau_* (t) }\,, \quad \ell_*\= \nu/u_*\,, \quad t_*\=
 \nu/ \tau _* (t)\ . \EE
 and denote the   normalized variables $ V^+\= {V}\big /
{u_*}$, $ K^+\= {K}\big /{u_*^2}$, $ z^+\= {z}/{\ell_*}$, $  t^+\=
{t}\big /{t_*}$. The friction velocity
 Reynolds number  is defined as  $
 \Rt(t)\= u_* \C Z(t)/\nu=\C Z(t)\big/ \ell_*= \C Z^+(t)
$, which is the width of TD-TBL in wall units. It should be noted
that a similar normalization is usually employed for the spatially
developing TBL (SD-TBL) except that the dependent  variables
 in that case are
functions of  the streamwise $x$ location instead of time.

The main advantage of the above normalization is that in wall units
the {\sl stationary} SD-TBL demonstrate universal behavior. We
assume  that the {\sl time-developing} TBL exhibits similar
universality in the limit of large times, as will be clarified
later. We will also show that the time-developing TBL has properties
very similar to that of the stationary TBL and its subregions  can
be classified in the same manner.\\

 \noindent \textbf{``Logarithmic accuracy" for
  ``asymptotically large times"}
  For ``asymptotically large times" we mean that $ \ln t^+\gg 1\ . $
We assume that in this limit there is a region in the TBL called the log-layer (with
$z^+ $ larger than upper boundary of the buffer sub-layer
$z^+\sb{buf}$), where
  the mean streamwise velocity profile is described by the \vK law:
  \BE \label{vK} V^+(z^+) =  \kappa^{-1} \ln z^+ + B\, , \quad z^+ >
   z^+\sb { buf}
  \approx 50 \ .\EE
  where $\kappa\approx 0.41$ is the \vK constant, and $B\approx 5.2
$.  We define the edge of the TD-TBL, $\C Z(t)$,
  as the normal distance from the wall where $V=V_\infty$.
  Thus, the \vK law at $\C Z^+(t)$ becomes: $
   V_\infty^+\= V^+(\C Z^+)=   \kappa^{-1} \ln \C Z^+ + B$.
  Moreover, in the limit $\ln t^+\gg 1$
    the width of TBL $\C Z(t)$ is large enough
  such that $\ln \C Z^+ \gg \k B \simeq 1$, thus,
  $ \ln \Rt\= \ln \C Z^+ \gg   1$.
  In the next section where we solve \Eq{MM},
  this inequality will allow us to neglect terms of order
   of unity with respect to terms of order $\ln \C Z^+$.
    We thus denote the accuracy of our results
     as the ``logarithmic accuracy". For example,
     with the logarithmic accuracy, ~\Eq{vK} for $V_\infty^+$ becomes 
  $ \k V_\infty^+= \ln \C Z^+  $. 

\begin{figure*}
\includegraphics[width=0.328   \textwidth]{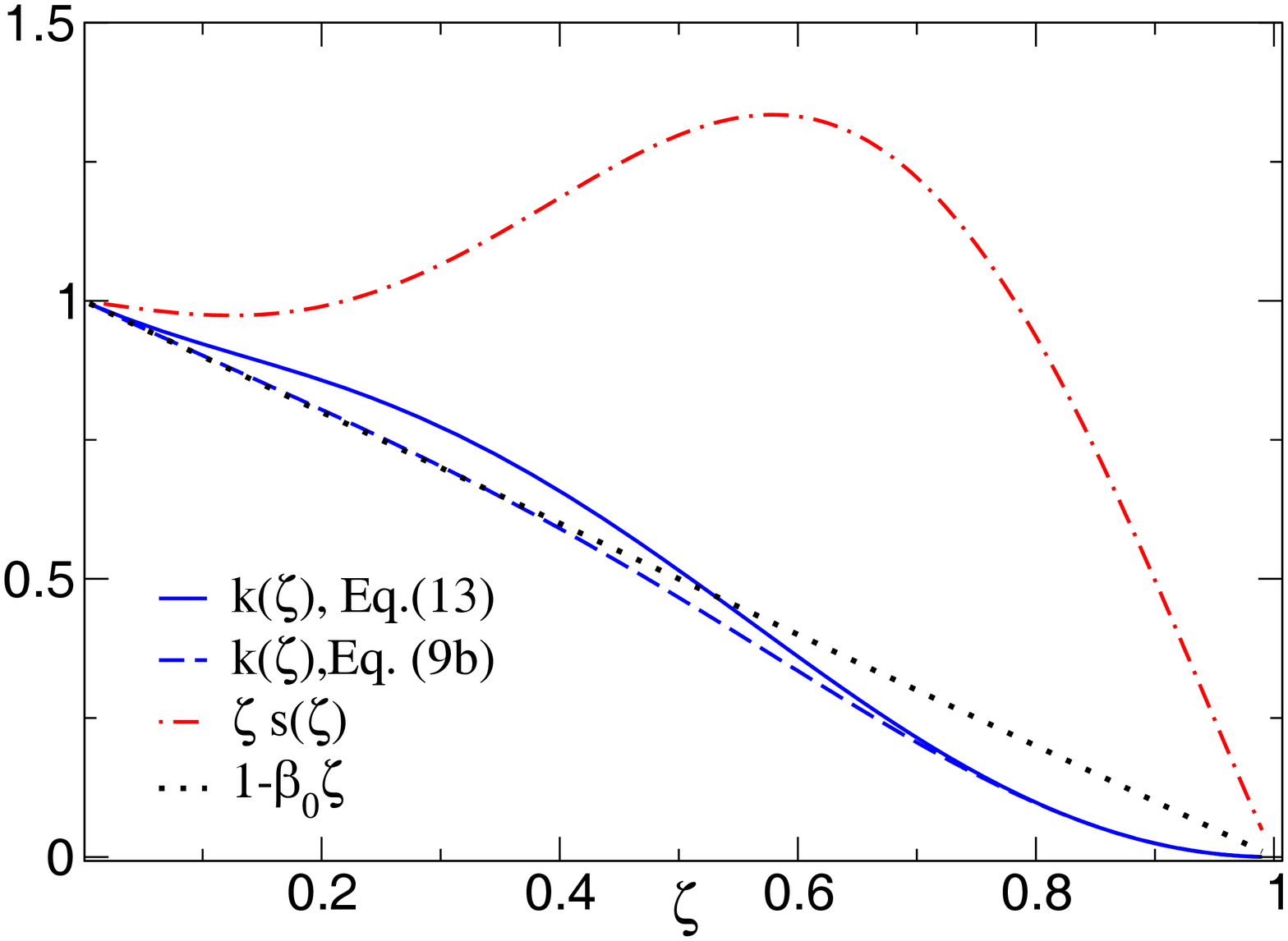}
   \includegraphics[width=0.328 \textwidth]{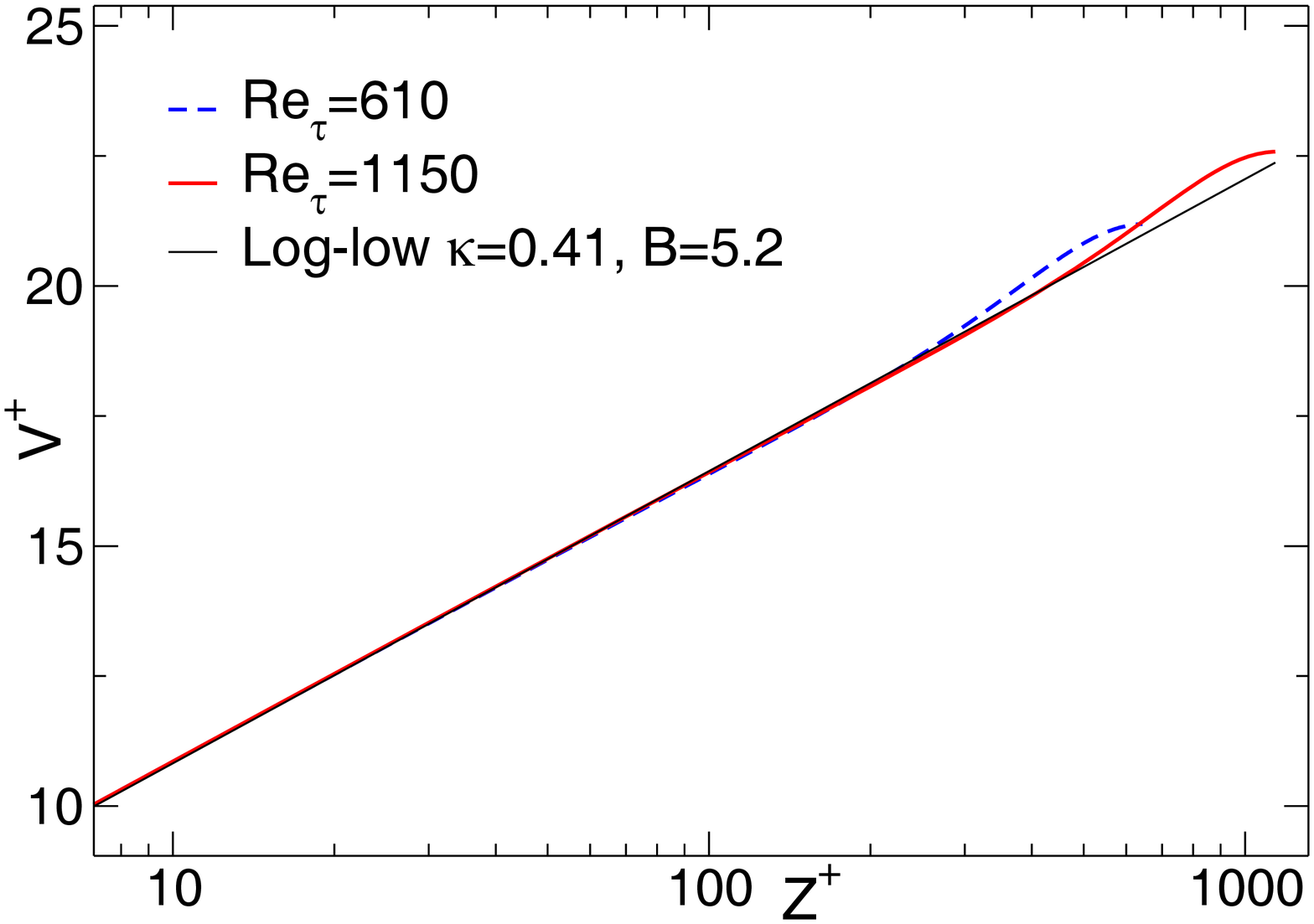}
    \includegraphics[width=0.33 \textwidth]{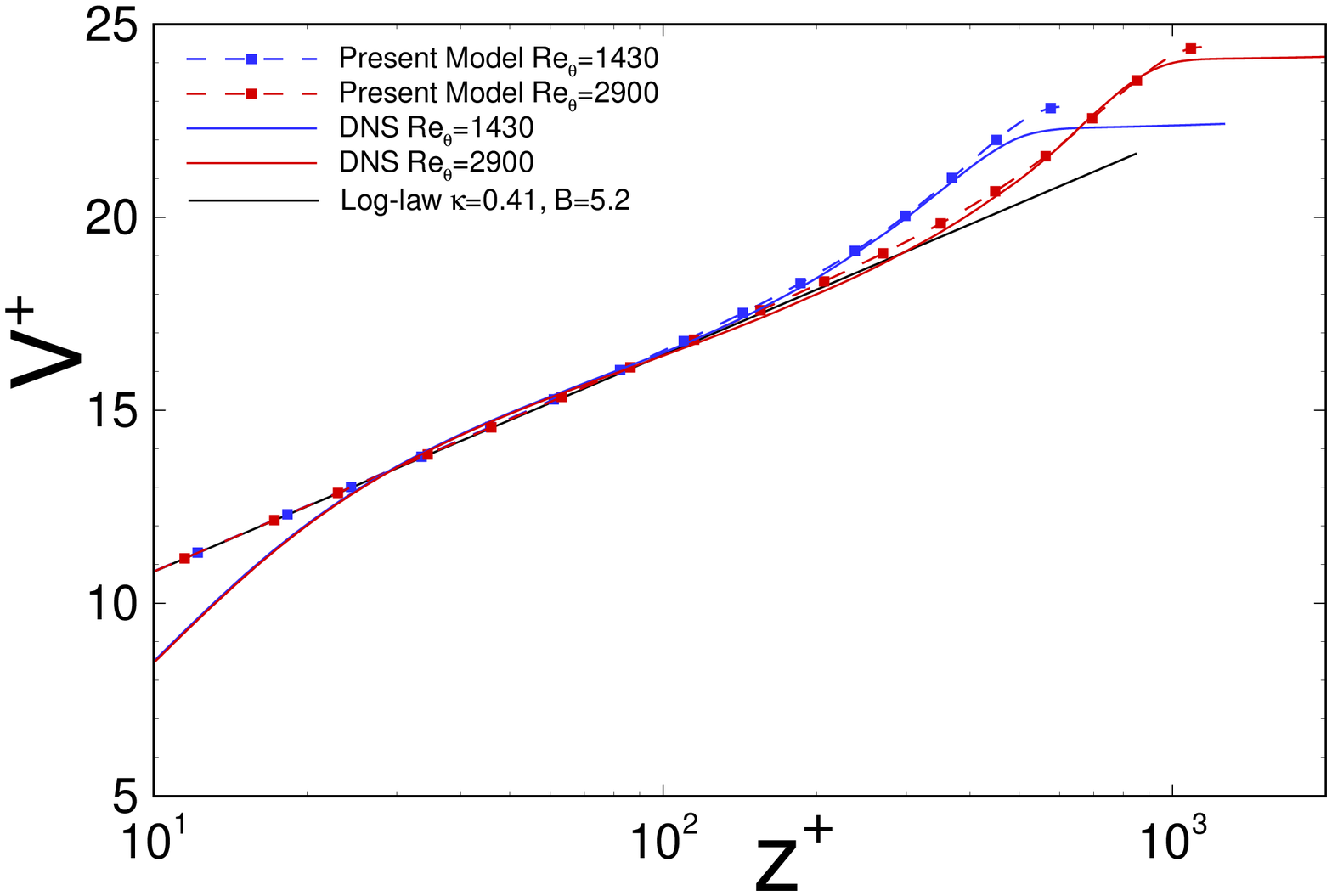}
\caption{\label{f:2} Color online. Left: Profile  of  $k(\zeta)$,
\Eq{rat} with fit constants, presented after~\Eq{con} is shown by
blue solid line, profile $\tilde k(\zeta)$, given by \Eq{b}, -- blue
dashed line and  small $ \zeta $ asymptotics $k(\zeta)=1- \b_0 \zeta
$ -- black dotted line. Profile of $\zeta  s(\zeta )$ is shown by
red dot-dashed line. Middle: Log-plot of mean velocity profile
$V^+(z^+)$ in the MM
 for $\Rt=610$  -- blue dashed line, $\Rt=1150$ -- red solid line
and  \vK log-law~\Eq{vK} -- thin black solid line. Right: Comparison
of the mean velocity profiles, given by improved Min-Model (dashed
lines) with DNS data (solid lines) for Re$_\theta=1430$ (giving
$\Rt\approx 610$) in blue, and  Re$_\theta=2900$ ($\Rt\approx
1150$)-- in red.}
\end{figure*}

 \noindent \textbf{Self-consistent  factorized solution}. In
order to reduce the system of partial differential \Eqs{MM} for the
unknown functions $S(z,t)$ and $K(z,t)$ of two variables $z$ and $t$
to a system of ordinary differential equations it is convenient to
introduce the following non-dimensional functions 
 \BSE\label{fac-eq}
  \BE \label{fac} k(\zeta)\= \frac{\tau^+(z,t)}{\tau_*(t)}\,, \
  s(\zeta)\= \frac{c^3 \C Z(t) S(z,t)}{ b\,  \tau_* (t)}\,, \EE
which {\sl in the turbulent region }$z^+>z^+\sb{buf}$ are  assumed
to be only
 functions of the ``outer variable" $\zeta=\zeta(t)\= z/\C Z(t)$.
Thus \Eq{MM1a} can be written as 
\BE\label{c} K(z,t)= \tau_*(t) k(\zeta)/c^2\ . \EE We assume also
that, the temporal growth of
the boundary layer is proportiona1 to the friction velocity,%
\BE \label{fac-c}  {d \C Z(t)}\big /{d t} = \a \sqrt{\tau_*(t)}
\,, \EE \ESE
with a dimensionless constant $\a$.

We now substitute $K(z,t)$, $S(z,t)$ and $\tau(z,t)$, expressed by
\Eqs{fac-eq} in terms of $k(\zeta)$, $s(\zeta)$ and $\tau_*(t)$
into the momentum and turbulence energy balance \Eqs{MM}. We also:
i) neglect the viscous contribution $\nu S $ in \Eq{MM1b} in the
region $z^+>z\sb{buf}^+$; ii) account only for the leading
contribution to $\p K/\p t$ (that originates from $d \C Z/ d t$)
and neglect the contribution of $d\tau_*/d t$. Similarly, in
\Eq{ME1} we account only for the leading contribution in $\ln
\Rt\gg 1$ terms. The result is two ordinary differential equations
for $k(\zeta)$ and $s(\zeta)$ with explicit expression for
$\tau_*(t)$:
\BSE\label{Fac-eq} \BEA\label{a}
 && -\a \, c\,  \zeta \frac{d  k  }{d  \zeta }+ \frac{b
 k  ^{3/2}}{\zeta }- d \frac{d}{d\zeta } \zeta  \sqrt {k }\,
 \frac{d
k }{d\zeta }=  b \, k   s  \,, \\ \label{b} && \kappa \frac{d\,k
}{d\, \zeta } + \a \, \zeta  \, s =0\,, \qquad\tau_*(t) =\Big[\frac{
\kappa V_\infty }{\ln (t/t_*)}\Big]^2 \ .~~~~\EEA\ESE 
In \Eq{a}, the RHS again represent the kinetic energy production,
and the three terms in the LHS describe temporal dependence, energy
dissipation and the diffusion, respectively.
 The
above system of equations for the functions $k(\zeta)$ and $s(\zeta)$
indicates that the factorization~\eq{fac} is consistent {\sl with the
logarithmic accuracy} of the momentum and turbulence energy balance
\Eqs{MM} in the sense that $k(\zeta)$ and $s(\zeta)$ depend only on
one variable $\zeta$.

The boundary conditions for these equations at the edge of TBL,
 $\zeta=1$, are:
\BSE\label{bc} \BE \label{bc1} \  s(1) = 0\,,\quad  \  k(1)=0\  .
\EE 
To formulate the boundary condition {\sl near} the wall, it is
noted that \Eqs{Fac-eq} are valid only for $z^+\ge
z^+\sb{buf}\approx 50$, which corresponds to $\zeta\ge
\zeta\sb{buf}\= z^+\sb{buf}/\Rt$. In this region the mean velocity
satisfies  the \vK law~\eq{vK} which gives $S^+(z^+\sb{buf})=1 /(
\k z^+\sb{buf})$. Noting that the full {Min-Model}~\eq{MM} leads
in the stationary case to $K^+(z^+\sb{buf})= c^{-2}$  and to
\Eq{vK} with $\k=c^3/b$,  we obtain with help of \Eq{fac}: 
\BE \label{bcB}
 \zeta \sb{buf} s(\zeta\sb{buf}) = 1\,, \    k(\zeta\sb{buf})
 = 1\  . \EE \ESE
Since $\zeta\sb{buf}\= z^+\sb{buf}/\Rt$, then we take the limit of
\Eq{bcB} as $\zeta\sb{buf}\to 0$ for asymptotically large times
when $\ln \Rt \gg 1$. It should be noted that boundary
conditions~\eq{bcB} are formulated {\sl near the wall} (in the
log-law region) but not {\sl at the wall}, in the viscous layer,
where $S^+\approx 1$.

With the boundary conditions~\eq{bcB} and \Eqs{fac} we obtain the
mean velocity $V(z,t)$ by simple integration:  
\BE \label{fac-d} V(z,t) = \sqrt{\tau_*(t)}\Big[B + \frac1\kappa
\int _{ {{\rm Re}_\tau^{-1}}}^{z/\C Z}s(\zeta ') d \zeta '
\Big]\,, \EE 
Here the lower limit of integration and term $B$ are chosen to
satisfy the \vK law, \Eq{fac-c}, in the turbulent log-law region
(for details, see Appendix).

  In order to solve analytically
  \Eq{fac-eq}, we now introduce a polynomial form of $k(\zeta)$:
\BE\label{rat} k(\zeta)= (1-\zeta )^2 (1+\b_1 \zeta  + \b_2 \zeta
^2 + \b_3 \zeta ^3 + \b_4 \zeta ^4 )^2\,, \EE 
which  satisfies the boundary conditions~\Eq{bc}.  Substituting
  (\ref{rat}) into (\ref{fac-eq}) and \eq{bc}, and neglecting terms
  of third-order and higher in the resulting equations give the
  five constants $\a\,, \beta_1\,,\dots \ \b_4$ as
\BEA\nn \b_1&=& 1 - \b_0/2\,, \quad \b_0\=\a/\k\,, \\
\nn\label{conB}  \b_2 &=&  \b_0(1+c\kappa / d )/2 -(2+\b_3+\b_4)\,, \\
\label{con}
\b_3 &=& 3- \b_0\, (1+13 c \kappa \b_0/6)/2 -2\b_4
\,,~~~~~~~~~~~~~~ \br \b_4&=&  -4+\b_0 \big[1+ c^4(c^4+ b d)/24+
371 c\kappa/ (108 d) \big]/2\ .\EEA 
Substituting $  \k=\a \int_0^1 \zeta  \, s(\zeta)\, d\zeta $
\tcblk{[}which follows from integrating \Eq{b} \tcblk {]} into
\Eq{conB},  the constants $\b_0\,,\dots\ \b_4$ can be expressed in
terms of the model's three parameters $b$, $c$ and $d$. For the
fixed values of $b=0.36$, $c=0.53$ and $d=0.07$, the resulting
values of $\beta_i$ are  $\b_0 \approx 0.996,\ \b_1\approx 0.502,
\ \b_2\approx 1.99,\ \b_3\approx -3,05,\ \b_4\approx 1.10$.

  Figure~\ref{f:2} (left) displays the profile of $k(\zeta)$ ( blue solid
line),
 given by \Eq{rat} and the above values of the
constants $\b_0\dots \b_4$. Substituting   this profile in the
energy balance~\Eq{a} we obtain the  profile of $\zeta s(\zeta)$,
shown by the red dot-dashed line. Substitution of the resulting
profile $\zeta s(\zeta)$ into \Eq{b} gives the profile of
$\~k(\zeta)$ shown in Fig.~\ref{f:2} (left) by the blue dashed line.
If the initial function $k(\zeta)$ gives the exact solution of the
balance \Eqs{fac-eq} then the functions $k(\zeta)$ and $\~k(\zeta)$
would  coincide.  \Fig{f:2} shows that these functions are
reasonably close. Thus, we conclude that the simple polynomial
form~\eq{rat} approximates the  solution of the balance \Eqs{fac-eq}
in the entire TBL, $0<\zeta <1$, with good accuracy.\\
\begin{figure*}
\includegraphics[width=0.325   \textwidth]{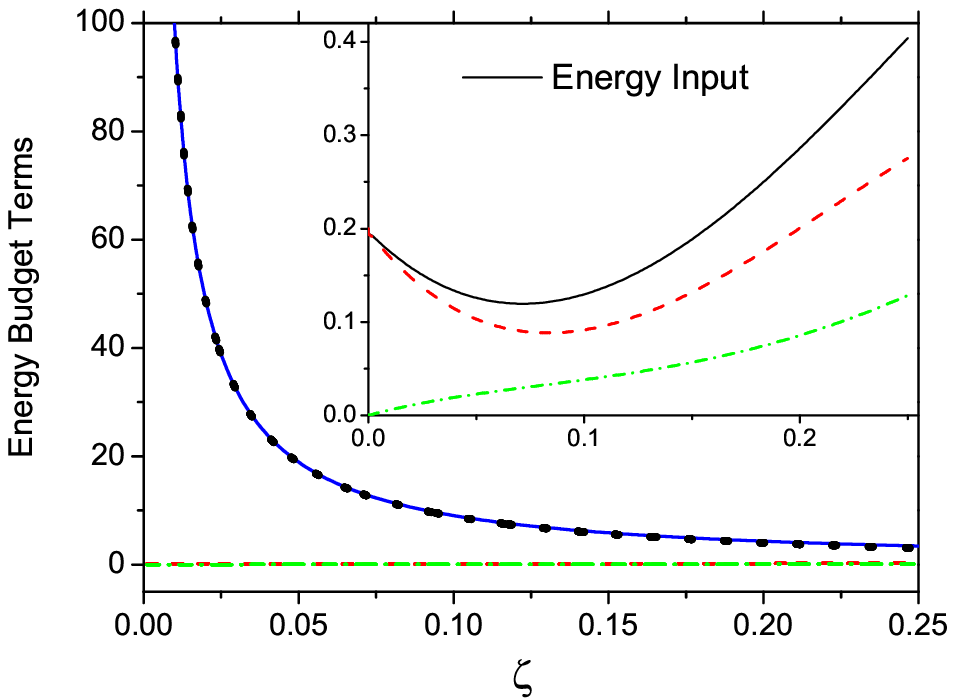}
  \includegraphics[width=0.33 \textwidth]{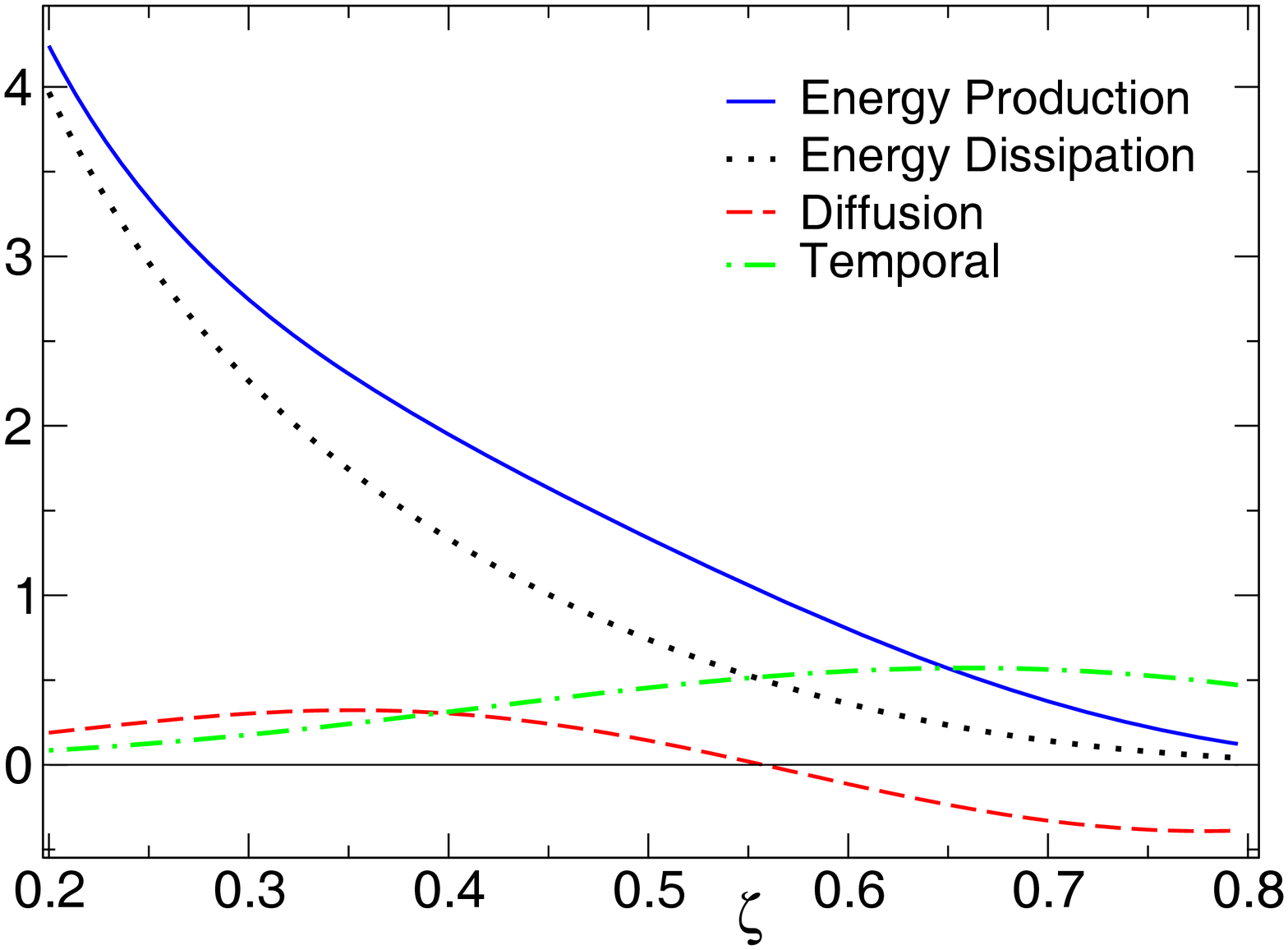}
 \includegraphics[width=0.33 \textwidth]{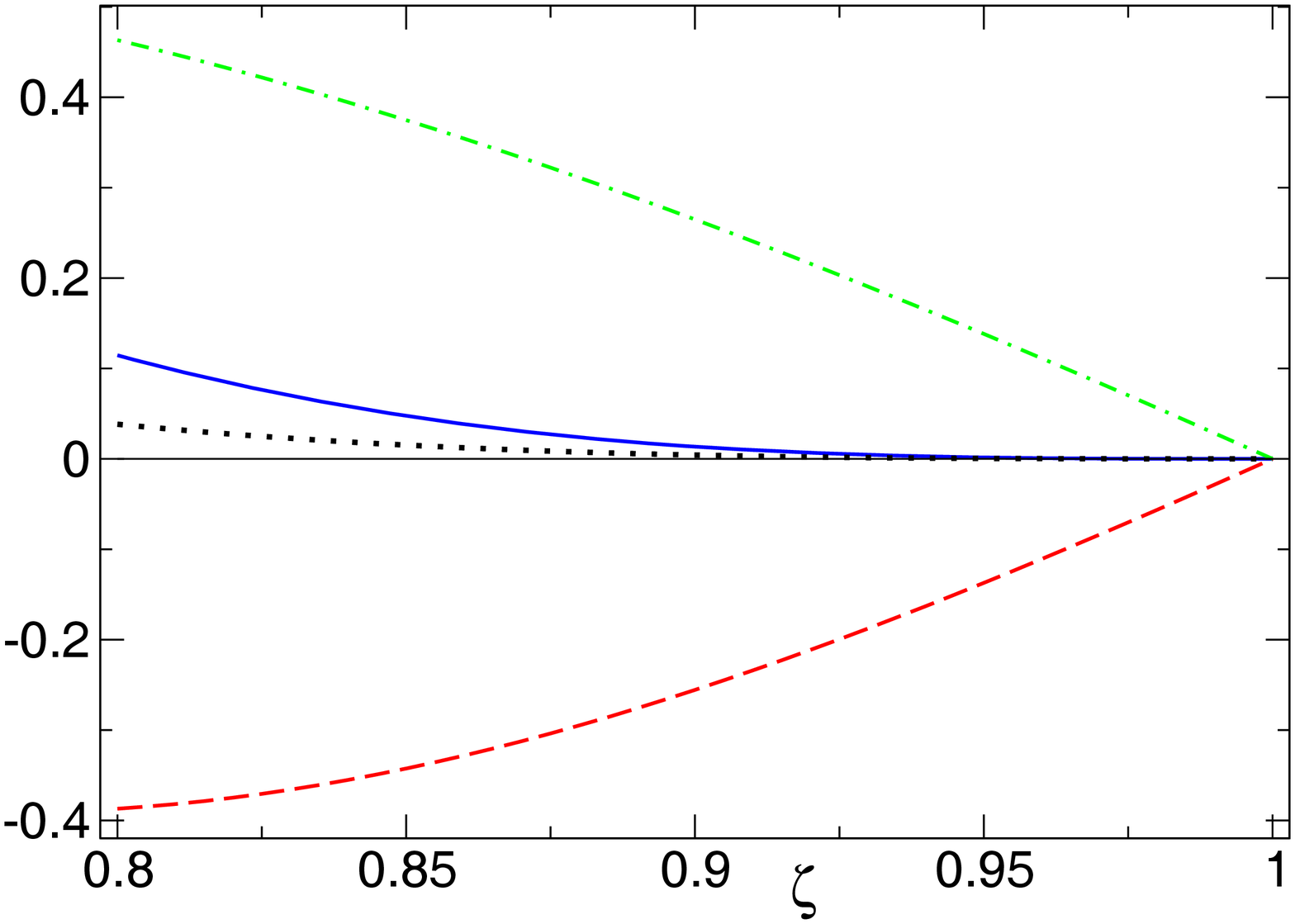}
\caption{\label{f:4} Color online. Energy balance in \Eq{a}. Energy
production term, $k(\zeta)\, s(\zeta)$ -- blue solid line, the
energy dissipation -- black dotted lines, the diffusion term red
dashed line, temporal term -- green dash-dotted lines and the energy
input (difference between the energy production and dissipation) --
black solid line in the insert of left panel. Different panels
display different region of TBL.  }
\end{figure*}

 \noindent\textbf{
 Comparison of  Min-Model  results with DNS}.
 In the present section, we compare the results of our
 Min-Model  for TD-TBL
with those obtained by DNS of SD-TBL~\cite{S1,S2,S3}. This
comparison is justified by the fact that TD-TBL and SD-TBL are
asymptotically equivalent in the limit of large $\Rt$~\cite{07LPR}.

\paragraph{ Mean streamwise velocity.} In our solution the unsteady  and
diffusion terms in the turbulence energy balance vanish near the wall,
for $ \zeta\to 0$. Accordingly in this region the mean streamwise
velocity $V(z,t)$ satisfies the \vK law~\eq{vK}, which gives [in our
normalization~\Eq{fac}] $s(\zeta)\to 1/ \zeta$.
 Indeed,in
Fig.~\ref{f:2} (left), $\zeta\, s(\zeta)$ become constant for small
$ \zeta $.

The figure also shows that the product $\zeta \, s(\zeta)$ in the
wide region $0 < \zeta<0.6$ exceeds unity, which is the level of
$\zeta s(\zeta)$ in the log-law layer. Accordingly, after
integration $s(\zeta)$ over $ \zeta $, the resulting mean velocity
profile $V_3(\zeta)$, shown in Fig.~\ref{f:2} (middle) exceeds the
log-law level.  Thus, the Min-Model clearly demonstrates the {\it
wake contribution}, described by Coles~\cite{wake} for stationary
channel flow.  The wake contribution is also seen in our DNS results
for the spatially- developing TBL.  However, the size of the wake,
given by Min-Model  Fig.~\ref{f:2}(Middle) is smaller than that in
DNS see Fig.~\ref{f:2} (Right).  This discrepancy originates from
the estimate of the outer scale of turbulence $\ell$. In the
Min-Model $\ell$ is estimated, following \vK, as the distance to the
solid wall $z$. This assumption is valid only for $z\ll \C Z$,
otherwise the scale $\ell$ is affected by the free upper boundary
and saturates.  Recently~\cite{LPR} the effect of saturation of
$\ell$ (at level of $0.3$ of the channel half-width) on the mean
velocity was studied in the channel flow. Accounting for the
saturation of $\ell$ in TD-TBL at the level $\approx \C Z/2$ (which
affects only the shear) improves the calculated velocity profile,
Fig.~\ref{f:2} (Right). The size of the wake, given by Improved
Min-Model compares well with that in DNS. This correction does not
effect the other results.

\begin{figure*}
\includegraphics[width=0.323 \textwidth]{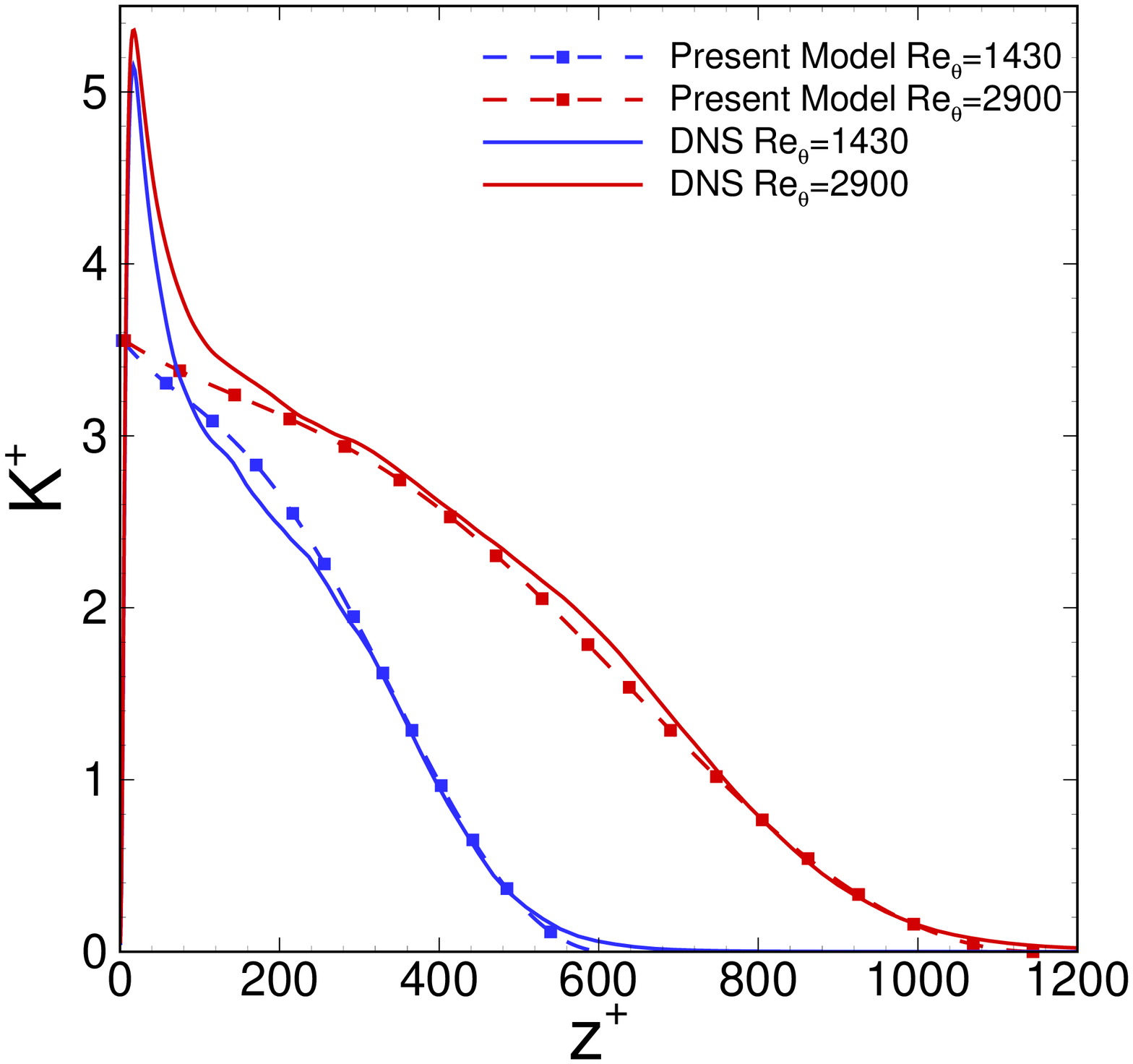}
 \includegraphics[width=0.323 \textwidth]{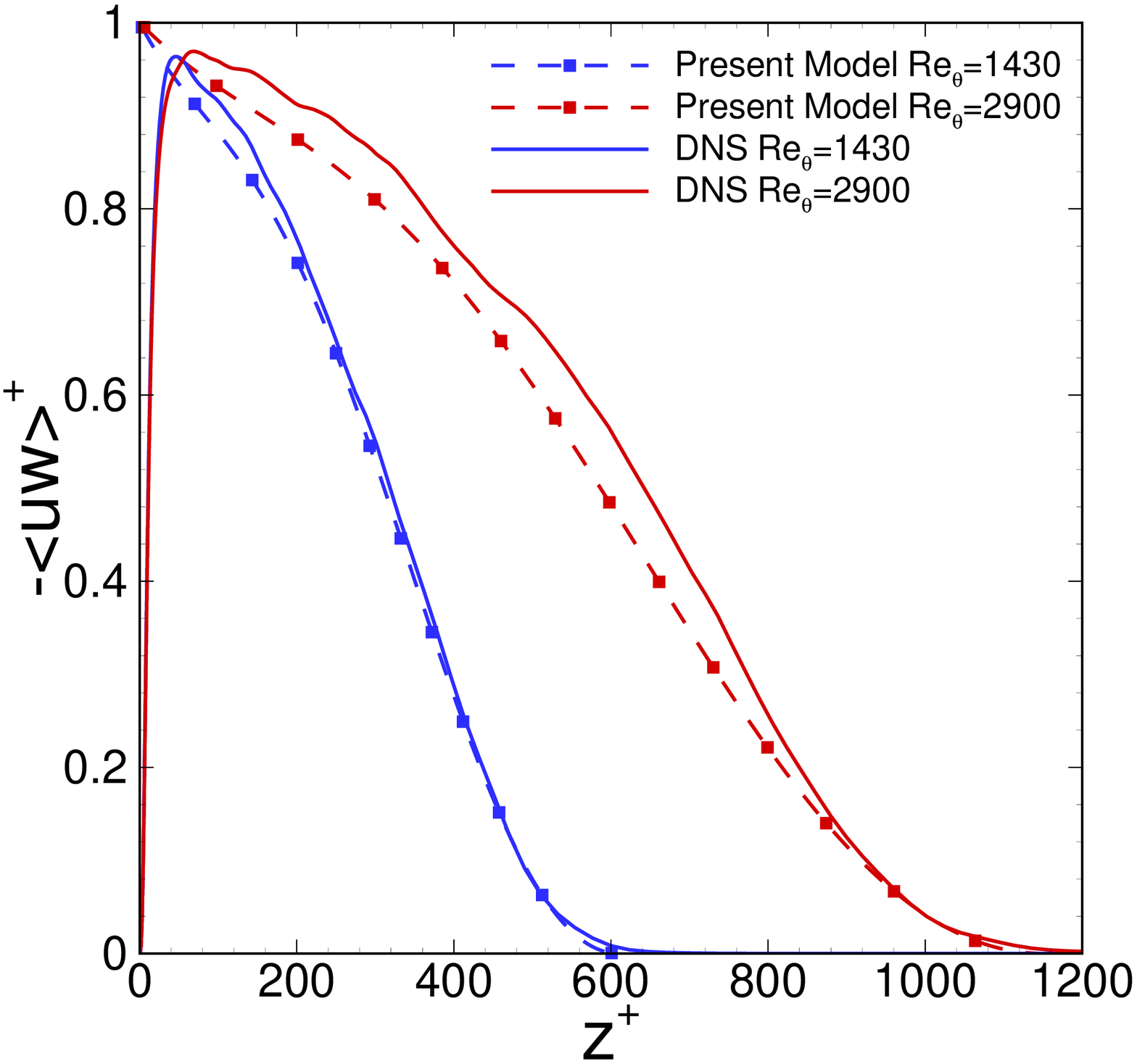}
 \includegraphics[width=0.323\textwidth]{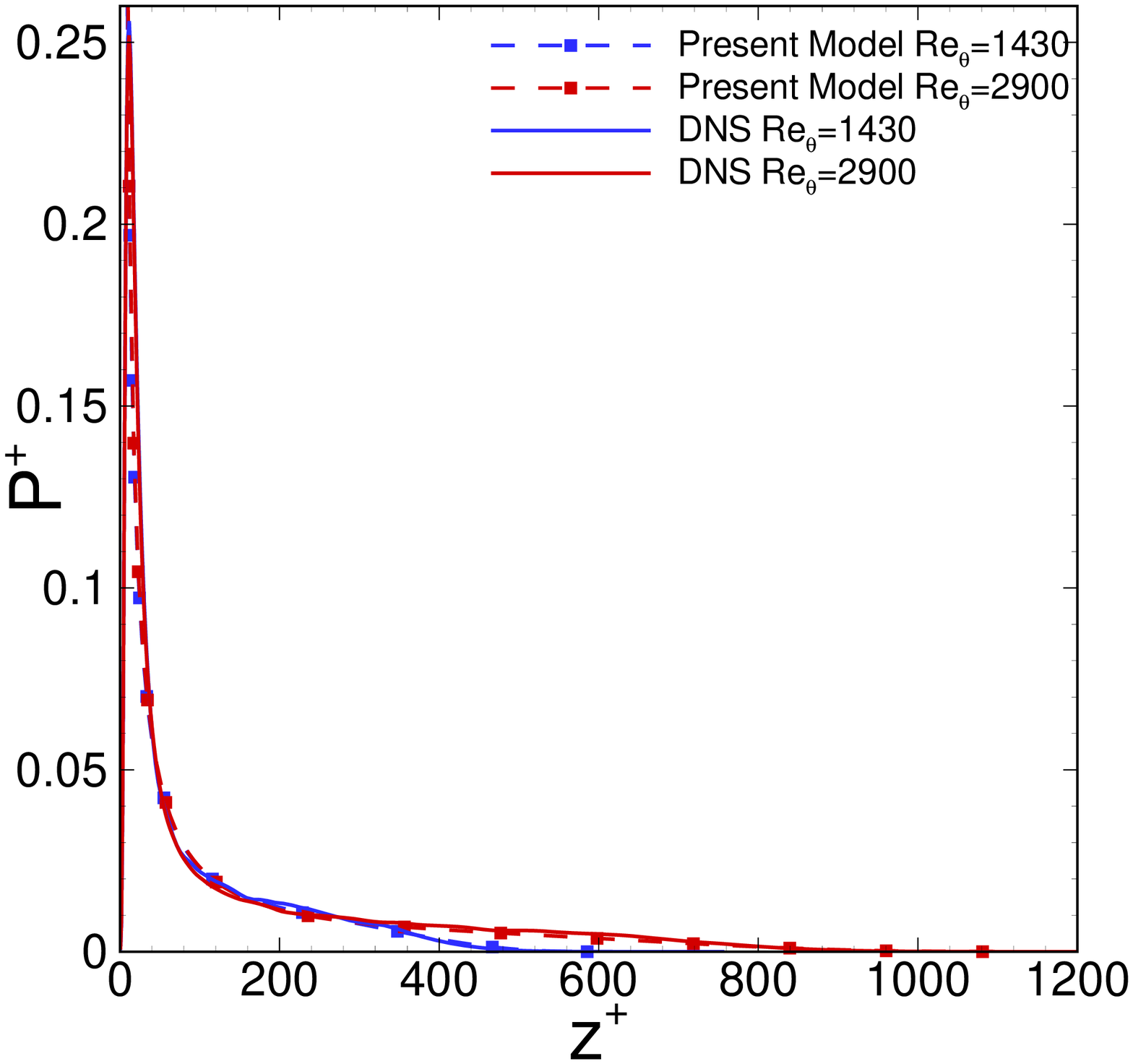}
\caption{\label{f:5}Color online.  Comparison of analytical
predictions (dashed lines) and DNS results (solid lines) for the
turbulence kinetic energy  (Left), Reynolds stress (Middle) and
production term (Right) . Blue lines corresponds to Re$_\theta$=1430
(giving $\Rt\approx 610$), red lines -- Re$_\theta$=2900
($\Rt\approx 1150$).}
\end{figure*}

\paragraph{Turbulence kinetic energy.}
 Figure~\ref{f:4} displays $\zeta $-dependence of various terms in the
energy balance.  One can see in the left panel  that close to the
wall, $\zeta < 0.25$, the energy production is well balanced by the
energy dissipation; the diffusion and unsteady terms play a minor
role.  They become relevant only for $ \zeta> 0.3$ as shown in the
middle panel. Thus the region, $ \zeta< 0.25$, can be considered as
``equilibrium" TBL with local spatial energy balance. For further
clarification, we plot in the insert   the difference between the
energy production and dissipation. The difference is denoted as the
{\sl energy input}. This is the part of the energy flux that is
required for the temporal development of TBL. Notice that the energy
input is finite, while the energy production and energy dissipation
themselves diverge as $1/ \zeta$  for $ \zeta\to 0$, as shown in
Fig.~\ref{f:4}, Left.

 The energy input is essential only in
the first half of the TBL, $ \zeta<\frac 12$.  As both energy
production and energy dissipation become smaller (see Fig.~\ref{f:4}
Middle and Right)  their difference vanishes. The energy input can
be totally neglected in the outer tenth of TBL, where $ \zeta>0.9$.
The only source of energy here is the turbulent diffusion, which
transports energy from the region $ \zeta \lesssim 0.5$ to the
region $ \zeta \gtrsim 0.5$. Thus, the turbulent diffusion
 leads  to increase the width of the TBL in time.
Figure~\ref{f:5}, Middle, shows good agreement between the
analytical (dashed lines) and DNS (solid lines) profiles of the
turbulence kinetic energy.  The observed discrepancy between the
analytical and DNS results for $z^+<50$ is expected since our
Min-Model was not designed to predict the buffer sub-layer flow.

\paragraph{Reynolds stress}  is one of the most important
characteristics of TBL, being responsible for the mechanical balance
in the turbulent log-law and outer regions. In stationary regimes,
the Reynolds stress is prescribed by the outer conditions,
maintaining the flow. For example, in pressure driven channel flow
(of half-width $L$) $\tau(z)\propto (1-z/L)$, while in the zero
pressure-gradient plane Couette flow $\tau(z)=$ const.  In
developing TBL, $\tau(z)$ is not known {\it a priory}, and thus the
Reynolds stress has to be determined self-consistently. Our model
[\Fig{f:2}(left),$k(\zeta)$] predicts  that in the first half of
TBL, near the wall $\tau(z)\approx [1-z/\C Z]$, as in the channel
flow, while in the last quarter of TBL, near the free boundary, it
decays quadratically, $\tau(z)\propto [1-z/\C Z]^2$. This prediction
is in a good agreement with the DNS data, \Fig{f:5} (Middle). The
production term is also well described by our model (see \Fig{f:5},
right panel)\\

\noindent\textbf{Summary}: We have presented a simple analytical
model (Min-Model) of the physics of time-developing TBL in a
Newtonian fluid.  The model is based on the exact equation for the
momentum flux and on the model equation for balance of turbulence
kinetic energy with the production, dissipation and turbulent
diffusion terms. The Min-Model results in a partial differential
equation for the kinetic energy, which, in the limit of large
evolution time was reduced to a relatively simple ordinary
differential equation. We obtained an asymptotic formula for the
time dependence of width of TBL and approximate analytical solutions
for the profiles of the mean velocity, the turbulent velocity
fluctuations and the Reynolds stress as a function of time and
distance from the wall. These profiles are in good agreement with
the DNS observations.  In future work we will demonstrate the
asymptotic equivalence of temporally- and spatially-developing
turbulent boundary layers by the relationship $x \Leftrightarrow
V_\infty t$ (here $x$ is the distance from the front edge and
$V_\infty$ is the free-stream velocity) and will clarify the
difference between these two regimes in pre-asymptotical region,
where $\ln t^+$ or $\ln x^+$ are not very large with respect of
unity.\\

  \noindent\textbf{Acknowledgements}. We thank Oleksii Rudenko for
fruitful discussions. We acknowledge the support of this research by
the US-Israel Binational Science Foundation.\\

  \noindent\textbf{Appendix.}
  Here we describe the derivation of \Eq{fac-d}. From the definition
  of mean shear \eq{defs}, the mean velocity is written as
  \be
  \label{eq:L1}
  V(z,t)= V(z\sb{buf},t)+ \int_{z\sb{buf}}^z S(z',t) d z' \ .
  \ee
  Substituting $S(z,t)$ from (\ref{fac}) into (\ref{eq:L1}) gives
  \be
  \label{eq:L2}
  V(z,t)= V(z\sb{buf},t) + \frac{\sqrt{\tau_*(t)}}{\kappa}
  \int_{\zeta \sb{buf}}^{\zeta}  s (\zeta ') d \zeta ' \ ,
  \ee\
  where $   \zeta \= z / \C Z(t)\,, \zeta\sb{buf}
  \= z\sb{buf} /\C  Z (t), \
$. The mean velocity $V(z\sb{buf})$ at the edge,
  $(z\sb{buf})$, of the log-law
region  is given by the \vK law as \be
  V(z\sb{buf})=
\sqrt{\tau_*}  \ [B+\ln z^+\sb{buf}]. \ee  Noting that $s (\zeta) =
1/\zeta $ for
  $\zeta\ll 1$,
    \Eq{eq:L2} becomes
\be
  \label{eq:L4}
  V(z,t)= \sqrt {\tau_*(t)} \Big [ B+
  \frac{1}{\kappa} \int_{\zeta_0}^\zeta  s ( \zeta ') d \zeta '\Big ]\ ,
  \ee
  where $\zeta_0= \ell_* / Z (t) = 1 / \Rt$.

\end{document}